%run once with writedefs and a second with input labeldefs.tmp
% Second revision, date 8/27/96
\input epsf

\def\INSERTCAP#1#2{\vbox{%
{\narrower\noindent%
\multiply\baselineskip by 3%
\divide\baselineskip by 4%
{\rm Table #1. }{\sl #2 \medskip}}
}}
%
%%%%%%%%%%%%%%%%%%%%%%%%%%%%%%%%%%%%%%%%%%%%%%%%%%%%%%%%%%%%%%%%%
% lanlmac is a hypertext utility.  If you don't have it,
% comment out \input lanlmac and uncomment the following line.
%%%%%%%%%%%%%%%%%%%%%%%%%%%%%%%%%%%%%%%%%%%%%%%%%%%%%%%%%%%%%%%%% 
\input lanlmac
%\input harvmac
% New Macros for Dstar
\def\vcb{$|V_{cb}|$}

%Some general use macros (mymacros.tex)
\def\OMIT#1{}

\def\spur{\raise.15ex\hbox{/}\kern-.57em }
%#1 removed from macro definition "spur"--Kelly, 09-03 

\def\ccdot{\hbox{\kern-.1em$\cdot$\kern-.1em}}
\def\frac#1#2{{#1\over#2}}

\def\larr#1{\raise1.5ex\hbox{$\leftarrow$}\mkern-16.5mu #1}

%
% definition for heavy mesons
%

%
% definitions put in by Glenn
\def\bcdecay{ $\Lambda_b \to \Lambda_c \ell \bar \nu$}

\def\lb{\Lambda_b}
\def\lc{\Lambda_c}
\def\mlb{ M_{\Lambda_b} }
\def\mlc{ M_{\Lambda_c} }
\def\eps{\epsilon}
\def\bra#1{\left\langle #1\right|}
\def\ket#1{\left| #1\right\rangle}

%
%definitions for HQET
%

\def\lsl{\spur {\kern0.1em l}}

%
%%%%
%%% some macros from my snowmass talk

%
%% macros for refs

\def\np#1#2#3{\NP{\bf B#1} (#2) #3}
\def\pl#1#2#3{\PL B {\bf #1} (#2) #3}
\def\prl#1#2#3{\PRL{\bf #1} (#2) #3}
\def\pr#1#2#3{\PR{\bf #1} (#2) #3}

\def\sjnp#1#2#3{\SJNP{\bf #1} (#2) #3}

\def\NP{{Nucl.\ Phys.\ }}
\def\PL{{Phys.\ Lett.\ }}
\def\PR{{Phys.\ Rev.\ }}

\def\PRL{{Phys.\ Rev.\ Lett.\ }}
\def\SJNP{{Sov.\ J. Nucl.\ Phys.\ }}
%
%
%
% macros for inserting figures
%
%
\def\INSERTFIG#1#2#3{\vbox{\vbox{\hfil\epsfbox{#1}\hfill}%
{\narrower\noindent%
\multiply\baselineskip by 3%
\divide\baselineskip by 4%
{\ninerm Figure #2. }{\ninesl #3 \medskip}}
}}%
%
%%  generic useful for ifs

%
\relax
%                   some macros
\def\w{{\omega}}

%
%\draft
\Title{\vbox{\hbox{UCSD/PTH 95-23}}}{\vbox{%
\centerline{Improved QCD Form Factor Constraints} 
\centerline{and $\Lambda_b \to \Lambda_c \ell \bar \nu$} }}
\centerline{C. Glenn Boyd\footnote{$^{\ast}$}{gboyd@ucsd.edu}
and Richard F. Lebed\footnote{$^{\ddagger}$}{rlebed@ucsd.edu}}
\bigskip\centerline{Department of Physics}
\centerline{University Of California, San Diego}
\centerline{La Jolla, California 92093-0319}
\vskip .3in
We construct model-independent parametrizations of the individual QCD
form factors relevant to $\Lambda_b \to \Lambda_c \ell \bar \nu$
decays. These results follow from dispersion relations and
analyticity, and incorporate an improvement in the technique that
reduces the number of necessary parameters. To describe most form
factors with 5\%--10\% accuracy over the entire kinematic range, three
parameters are necessary, one of which is its normalization at zero
recoil.  We also apply the improvement to meson decays, and find,
using the heavy quark form factor normalization, that almost every
$\bar B \to D \ell \bar \nu$ and $\bar B \to D^* \ell \bar \nu$ form
factor is well-described by a single-variable parametrization.  $\bar
B \to \pi \ell \nu$ requires a total of only 3 to 5 parameters,
depending on the desired accuracy.
\Date{Final Version: October, 1996} 
\nref\iw{N. Isgur and M.B. Wise, \pl{232}{1989}{113} and
\pl{237}{1990}{527}.}
\nref\eh{E. Eichten and B. Hill, \pl{234}{1990}{511}.}
\nref\vs{M. B. Voloshin and M. A. Shifman, Yad.\ Fiz.\ {\bf 47},
(1988) 801 [\sjnp{47}{1988}{511}].}
\nref\iwbary{N. Isgur and M. B. Wise, \np{348}{1991}{276}.}
\nref\luke{M. E. Luke, \pl{252}{1990}{447}.}
\nref\ggw{H. Georgi, B. Grinstein, and M. B. Wise,
\pl{252}{1990}{456}.}
\nref\bdaleph{D. Buskulic {\it et al.\/} (ALEPH Collaboration),
\pl{359}{1995}{236}.}
\nref\vivek{V. Sharma, private communication.}
\nref\cern{ALEPH Collaboration, contribution to the International
Europhysics Conference on High Energy Physics, Brussels, Belgium, July
1995, EPS0753\semi DELPHI Collaboration, CERN-PPE/95-54\semi OPAL
Collaboration, CERN-PPE/95-051.}
\nref\cdf{CDF Collaboration, HyperText document
http://www-cdf.fnal.gov/phy\-sics/new/bot\-tom/cdf3395/cdf3395.html.}
\nref\hist{N.N. Meiman, Sov.\ Phys.\ JETP {\bf 17} (1963) 830\semi
S. Okubo, \pr{D3}{1971}{2807}\semi
S. Okubo and I. Fushih, \pr{D4}{1971}{2020}\semi
V. Singh and A.K. Raina, Fortschritte der Physik {\bf 27} (1979)
561.}
\nref\bmd{C. Bourrely, B. Machet, and E. de Rafael,
\np{189}{1981}{157}.}
\nref\drtone{E. de Rafael and J. Taron, \pl{282}{1992}{215}.}
\nref\spoilers{E. Carlson, J. Milana, N. Isgur, T. Mannel, and
W. Roberts, \pl{299}{1993}{133}\semi
A. Falk, M. Luke, and M. B. Wise, \pl{299}{1993}{123}\semi
B. Grinstein and P. Mende, \pl{299}{1993}{127}\semi
C. Dominguez, J. K\"{o}rner, and D. Pirjol, \pl{301}{1993}{373}.}
\nref\rtc{
E. de Rafael and J. Taron, \pr{D50}{1994}{373}\semi
I. Caprini, Z. Phys.\ C {\bf 61} (1994) 651, \pl{339}{1994}{187}.}
\nref\dec{C.G. Boyd, B. Grinstein, and R.F. Lebed,
\prl{74}{1995}{4603}.}
\nref\bglapr{C.G. Boyd, B. Grinstein, and R.F. Lebed,
\pl{353}{1995}{306}.}
\nref\bglaug{C.G. Boyd, B. Grinstein, and R.F. Lebed,
\np{461}{1996}{493}.}
\nref\lellouch{L. Lellouch, Marseille preprint CPT-95/P.3236
[hep-ph/9509358] (unpublished).}
\nref\Koerner{J. G. K\"{o}rner and M. Kr\"{a}mer,
\pl{275}{1992}{495}.}
\nref\blasref{P. Duren, {\it Theory of $H^p$ Spaces}, Academic Press,
New York, 1970.}
\nref\alephbc{ALEPH Collaboration, Report No.\ EPS0407, contribution
to the International Europhysics Conference on High Energy Physics,
Brussels, Belgium, July, 1995.}
\nref\klt{V.V. Kisilev, A.K. Likhoded, and A.V. Tkabladze,
\pr{D51}{1995}{3613}.}
\nref\eq{E.J. Eichten and C. Quigg, \pr{D49}{1994}{5845}.}
\nref\ck{Y.-Q. Chen and Y.-P. Kuang, \pr{D46}{1992}{1165}.}
\nref\quigg{Chris Quigg, private communication.}
\nref\bjbd{J.D. Bjorken, in {\it Proceedings of the 4th Recontres de
Physique de la Vall\`{e}e d'Aoste}, La Thuille, Italy, 1990, ed.\
M. Greco (Editions Fronti\`{e}rs, Gif-Sur-Yvette, France, 1990)\semi
N. Isgur and M.B. Wise, \pr{D43}{1991}{819}\semi
J.D. Bjorken, I. Dunietz, and J. Taron, \np{371}{1992}{111}.}
\nref\bgm{C.G. Boyd, B. Grinstein, and A.V. Manohar, U.C. San Diego
preprint No.\ UCSD/PTH 95-19 [hep-ph/9511233] (to appear in Phys.\
Rev.\ D).}
\nref\models{M. Wirbel, B. Stech, and M. Bauer, Zeit.\ Phys.\ {\bf
C29} (1985) 637\semi
N. Isgur and D. Scora, \pr{D52}{1995}{2783}.}
\newsec{Introduction}
	Considerable theoretical and experimental attention has been
devoted to the extraction of the Cabibbo--Kobayashi-Maskawa element
\vcb\ from both exclusive and inclusive semileptonic decays of the $B$
meson.  Because of the key role this parameter plays in the
investigation of rare decays and CP violation in the third quark
generation, it is important to make as many independent determinations
of \vcb\ as possible.  The baryonic semileptonic decay \bcdecay\
provides an opportunity to extract \vcb\ in a fashion as theoretically
clean as in $\bar B \to D \ell \bar \nu$ or $\bar B \to D^* \ell \bar
\nu$, because heavy quark symmetry\refs{\iw{--}\vs} predicts for both
cases a single universal form factor in the heavy quark
limit\refs\iwbary, including the normalization of $1 + {\cal O}
(1/m_c^2)$ at zero recoil\refs{\luke,\ggw}.

	The prospect for experimentally determining the $q^2$
dependence of \bcdecay\ form factors is promising, because the method
applied by ALEPH\refs\bdaleph\ to $\bar B \to D^* \ell \bar \nu$
should work equally well\refs\vivek\ for baryons. The cumulative data
sets of ALEPH, DELPHI, and OPAL\refs\cern\ contain about 250
$\Lambda_b$ semileptonic events; an additional 200 events have been
observed at CDF\refs\cdf. This raises the possibility of extracting
\vcb\ from baryon decays.

	To fully exploit this possibility, a model-independent
parametrization of the \bcdecay\ form factors is desirable. This is
important because the kinematic rate vanishes at zero recoil, so an
extrapolation of the data based on some parametrization is required.

	In this paper we apply the parametrization imposed by QCD on
the physical form factors.  The requirement of compatibility with QCD
is obtained through the application of dispersion relations based on
the analyticity properties of form factors as functions of the
momentum transfer variable.  The method of extracting information on
amplitudes by this form of complex analysis is quite old\refs\hist,
and has been applied to the study of semileptonic decays of light
mesons in a more contemporary language in Ref.~\refs\bmd.  Its
application to heavy quark systems has received much attention in more
recent years\refs{\drtone{--}\lellouch}.

	In general, these applications say little about the overall
normalization of form factors, but encode a great deal of information
about the shape.  The expression of this information manifests itself
in a simple parametrization\refs{\bglapr,\bglaug}\ that spans the
functional space of form factors allowed by QCD dispersion relations.
In this work we apply the parametrization, supplemented by a new
development presented here, to the decay \bcdecay.

	The technical development introduced for \bcdecay\ also has
important implications for form factors in meson decays such as $\bar
B \to D^{(*)} \ell \bar \nu$ and $\bar B \to \pi \ell \bar \nu$. It
allows one to obtain form factor parametrizations with smaller
uncertainties and fewer parameters than those discussed in earlier
works\refs{\bglapr,\bglaug}.

	The paper is organized as follows: In Sec.\ 2 we outline the
derivation of form factor bounds from QCD dispersion relations and
introduce the technical improvement that strengthens our constraints.
We construct the form factor parametrization that obeys these bounds,
and discuss the inclusion of heavy resonances, in Sec.\ 3.  In Sec.\ 4
we discuss the numerics of the heavy resonance masses and obtain
information on the quality of parametrizations for \bcdecay\ form
factors.  Section 5 describes the implications of our technical
improvement to meson decays.  We conclude in Sec.\ 6.

\newsec{Dispersion Relation Inequalities}

The QCD matrix elements governing the semileptonic decay \bcdecay\ may
be expressed in terms of form factors defined by
\eqn\fdefs{\eqalign{
  \bra{\Lambda_c(p')} V^\mu \ket{\Lambda_b(p)} =&
 \bar u_c(p') [ F_1\gamma^\mu+F_2v^\mu+F_3v^{\prime \mu} ] u_b(p), \cr
\bra{\Lambda_c(p')} A^\mu \ket{\Lambda_b(p)} =&
 \bar u_c(p') [ G_1\gamma^\mu+G_2v^\mu+G_3v^{\prime \mu} ]
\gamma_5 u_b(p),
\cr }}
where $v = p/\mlb $ and $v' = p'/\mlc$ are meson velocities, $V^\mu =
\bar c \gamma^\mu b$ and $A^\mu = \bar c \gamma^\mu \gamma_5 b$ are
polar and axial vector flavor-changing currents, and the form factors
are functions of the momentum transfer $q^2 = (p - p')^2$. In terms of
these form factors, the differential decay width for \bcdecay\ with a
massless charged lepton $\ell$ is
\eqn\rate{
{d\Gamma \over dq^2} = {|V_{cb}|^2 G_F^2 (k^2 q^2)^{\frac12} \over 96
\pi^3 M^3} \Bigl\{ (Q_-) \left[ 2 q^2 |F_1|^2 + M^2 |H_V|^2 \right] +
(Q_+) \left[ 2 q^2 |G_1|^2 + M^2 |H_A|^2 \right] \Bigr\} , }
where the terms proportional to $|F_1|^2$ and $|G_1|^2$ alone give the
partial widths for transversely polarized intermediate $W$
bosons\refs\Koerner, whereas
\eqn\Hedefs{
\eqalign{ H_V(q^2) &= {1 \over M} \left[ (M + m) F_1 + {Q_+ \over 2}
\left( {F_2 \over M} + {F_3 \over m} \right) \right] , \cr H_A(q^2) &=
{1 \over M} \left[ (M- m) G_1 - {Q_- \over 2} \left( {G_2 \over M} +
{G_3 \over m} \right) \right] , \cr }}
are the partial wave amplitudes for a longitudinally polarized $W$'s,
with
\eqn\Qdefs{
Q_{\pm} = (M \pm m)^2 - q^2 .
}
The factor 
\eqn\kdef{
k^2 = {M^2 \over q^2} {\bf p}_c^2 = {1\over 4 q^2} Q_+ Q_-
}
is an invariant kinematic quantity related to the three-momentum ${\bf
p}_c$ of the $\Lambda_c$ in the rest frame of the decaying
$\Lambda_b$, and $M,m = \mlb, \mlc$, respectively.  The form factor
combinations $\left({F_2 \over M} - {F_3 \over m}\right)$ and
$\left({G_2 \over M} - {G_3 \over m}\right)$, which appear with the
Lorentz structure $q^\mu$, give contributions to the rate proportional
to the lepton mass squared $m_\ell^2$, and are consequently mainly of
interest for constraining models.

	We begin our derivation of constraints from dispersion
relations following the well-known methods developed by the authors
listed in Refs.~\refs{\hist,\bmd}.  In QCD, the two-point function of
a flavor-changing current $J = V, A,$ or $V-A$,
\eqn\twopntfnctn{
\Pi^{\mu \nu}_J (q) = (q^\mu q^\nu-q^2g^{\mu\nu})\Pi_J^T(q^2) +
g^{\mu\nu}\Pi_J^L(q^2) \equiv {i \int d^4\!x \, e^{iqx}\vev{0|{\rm T}
J^\mu(x) J^{\dagger\nu}(0)|0}, }}
can be rendered finite by making one subtraction, leading to the
dispersion relations
\eqn\disper{
\chi^{T,L}_J (q^2)\equiv\left.{{\partial\Pi_J^{T,L}}
\over{\partial q^2}}\right.=
{1\over\pi}\int_0^\infty dt \, {{{\rm
Im}\,\Pi_J^{T,L}(t)}\over{(t-q^2)^2}}
.}
The functions $\chi^{T,L}_J(q^2)$ may be computed reliably in
perturbative QCD for values of $q^2$ far from the kinematic region
where the current $J$ can create resonances: specifically,
$(m_b+m_c)\Lambda_{\rm QCD} \ll (m_b+m_c)^2 - q^2$.  For resonances
containing a heavy quark, it is sufficient to take $q^2=0$.  At one
loop, $\chi_V^L = 3.7 \cdot 10^{-3}$ and $\chi_A^L = 2.2 \cdot
10^{-2}$. We also make use of the quantity
$\chi^{\vphantom{\dagger}}_J \equiv \chi^T_J(0) - \frac12
\frac\partial{\partial q^2} \chi^L_J(0)$, corresponding to the
combination of $\Pi^T_J$ and $\Pi^L_J$ that gives $\Pi^{ii}_J$ at
$q^2=0$.  The full one-loop expressions for $q^2=0$ are
\eqn\chis{\eqalign{
\chi^{\vphantom{\dagger}}_V(u) = \chi^{\vphantom{\dagger}}_A(-u)
   = & \, {1 \over 32 \pi^2 m_b^2 (1 - u^2)^5 } \cr & \times [ (1 -
   u^2)(3+4 u- 21 u^2 + 40 u^3 - 21 u^4 + 4 u^5 + 3 u^6) \cr & \; + 12
   u^3 (2 - 3 u + 2 u^2)\ln u^2 ], \cr
\chi^L_V (u) = \chi^L_A (-u) =  & {1 \over 8\pi^2 (1 - u^2)^3} \left[
(1 - u^2) (1 + u + u^2) (1 - 4 u + u^2) - 6 u^3 \ln u^2 \right] ,
}}
where $u = m_c/m_b$ is the ratio of quark masses.  For bottom and
charm quark masses of $m_b = 4.5$ GeV and $m_c = 1.5$ GeV, the
one-loop values of $M^2 \chi_V$ and $M^2 \chi_A$ are $1.5 \cdot
10^{-2}$ and $9.0 \cdot 10^{-3}$, respectively.  One should keep in
mind that ${\cal O} (\alpha_s)$ corrections to the bounds derived
below enter as corrections to these values.

	The absorptive part ${\rm Im}\,\Pi_J^{\mu \nu}(q^2)$ is
obtained by inserting on-shell states between the two currents on the
right-hand side of Eq.~\twopntfnctn.  For any four-vector $n_\mu$, the
quantity $n_\mu^*\Pi_J^{\mu \nu} n_\nu$ is a sum of positive-definite
terms, so one can obtain strict inequalities by concentrating on the
term with the two-particle intermediate state $\bar \Lambda_b
\Lambda_c$.  The contribution of $\bar \Lambda_b \Lambda_c$ pairs to
the right-hand side of \disper\ enters as
\eqn\Xineq{\eqalign{
n_\mu^*{\rm Im}\,\Pi^{\mu\nu}_{V-A} n_\nu  
 &\geq \theta(q^2-(M+m)^2)
{ \sqrt{k^2} \over 8 \pi^2 \sqrt{q^2} } \int d\Omega_p \,
\Bigl\{ M m X_1 (n^*\cdot n)  \cr
    &- \frac{m}M X_2 |p\cdot n|^2 
- X_3 i \epsilon^{\mu\nu\alpha\beta} n_\mu^*n_\nu p_\alpha q_\beta
- \frac{M}m X_4 |(q - p) \cdot n|^2 
\cr &+ X_5 (p\cdot n^*)  (q - p)\cdot n 
+ X_5^* (p\cdot n)  (q - p) \cdot n^* \Bigr\},\cr
}}
where
\eqn\Xdefs{\eqalign{
X_1 & = \left( \w - 1 \right)|F_1|^2 + \left( 1 + \w \right)
|G_1|^2,\cr
X_2 &= \left(1 + \w\right) |F_2|^2 + (F_1^* F_2 + F_1 F_2^*)
+\left(\w-1\right) |G_2|^2 + (G_1^* G_2 + G_1 G_2^*), \cr
X_3 &= (F_1^* G_1 + F_1 G_1^*),\cr
X_4 &= \left(1 + \w\right) |F_3|^2 + (F_1^* F_3 + F_1 F_3^*) +
\left(\w-1\right) |G_3|^2 - (G_1^* G_3 + G_1 G_3^*), \cr
X_5 &= \left(1 + \w \right) F_2^* F_3 + \left(\w -1 \right)G_2^* G_3 +
|F_1|^2 + F_1 F_2^* + F_1^* F_3 + |G_1|^2 - G_1 G_2^* + G_1^* G_3,
}}
with $\w \equiv (M^2 + m^2 - q^2)/(2 M m)$, and the integration is
over all directions of the momentum vector $p$.  For massless leptons
the differential width Eq.~\rate\ is proportional to the space-space
components ${\rm Im}\,\Pi^{ii}_J$ of the two-point function, so the
choice $n = (0,\hat {\bf n})$ leads to inequalities on the physically
interesting form factors $F_1$, $G_1$, $H_V$, and $H_A$. For
completeness, we also consider $n = (1,{\bf 0})$, which leads to
inequalities on the combinations
\eqn\fgzero{
\eqalign{ F_0 &= {1 \over M} \left[ (M-m) F_1 + {1 \over 2M} (q^2 +
M^2 -m^2) F_2 - {1 \over 2m} (q^2 - M^2 +m^2) F_3 \right] \cr G_0 &=
{1 \over M} \left[ (M+m) G_1 - {1 \over 2M} (q^2 + M^2 -m^2) G_2 + {1
\over 2m} (q^2 - M^2 +m^2) G_3 \right] . \cr }}
However, since $n_\mu$ can be an arbitrary $q^2$-dependent
four-vector, there is considerable freedom to constrain other
combinations of form factors.

The analysis simplifies when Eq.~\disper\ is written in terms of the
conformally transformed variable $z$ defined by
\eqn\zdef{
{1+z \over 1-z} = \sqrt{(M+m)^2 -q^2 \over 4 N M m }
= \sqrt{(1 + \w) \over 2 N}.
}
This expression differs from that previously used in the literature by
the inclusion of the factor $N$; the two agree when $N=1$.  Upon
choosing the principal branch of the square root in this expression,
the change of variables $q^2 \to z$ maps the two sides of the cut $q^2
> (M+m)^2 $ to the unit circle $|z|=1$, with the rest of the $q^2$
plane mapped to the interior of the unit circle (Fig.\ 1).  In
particular, the real values $-\infty < q^2 \leq (M-m)^2 -4 (N-1) M m$
and $(M-m)^2 -4 (N-1) M m\leq q^2 < (M+m)^2$ are mapped to the real
axis, $1 > z \geq 0$ and $0 \geq z > -1$ respectively.  Physically,
this means that the kinematic region relevant to the process {\it
vacuum} $\to \overline{M} m$, $q^2 \geq (M+m)^2$, lies on the unit
circle, while the region for semileptonic decay $M \to m \ell \bar
\nu$ for massless $\ell$, $0 \leq q^2 \leq (M-m)^2$, lies inside the
circle on the real $z$-axis.  Specifically,
\eqn\zbounds{\eqalign{
z_{\rm min} & = - \left( \frac{\sqrt{N}-1}{\sqrt{N}+1} \right) , \cr
z_{\rm max} & = \frac{(1-\sqrt{r})^2 -
2\sqrt{r}(\sqrt{N}-1)}{(1+\sqrt{r})^2
+ 2\sqrt{r}(\sqrt{N}-1)} ,
}}
\INSERTFIG{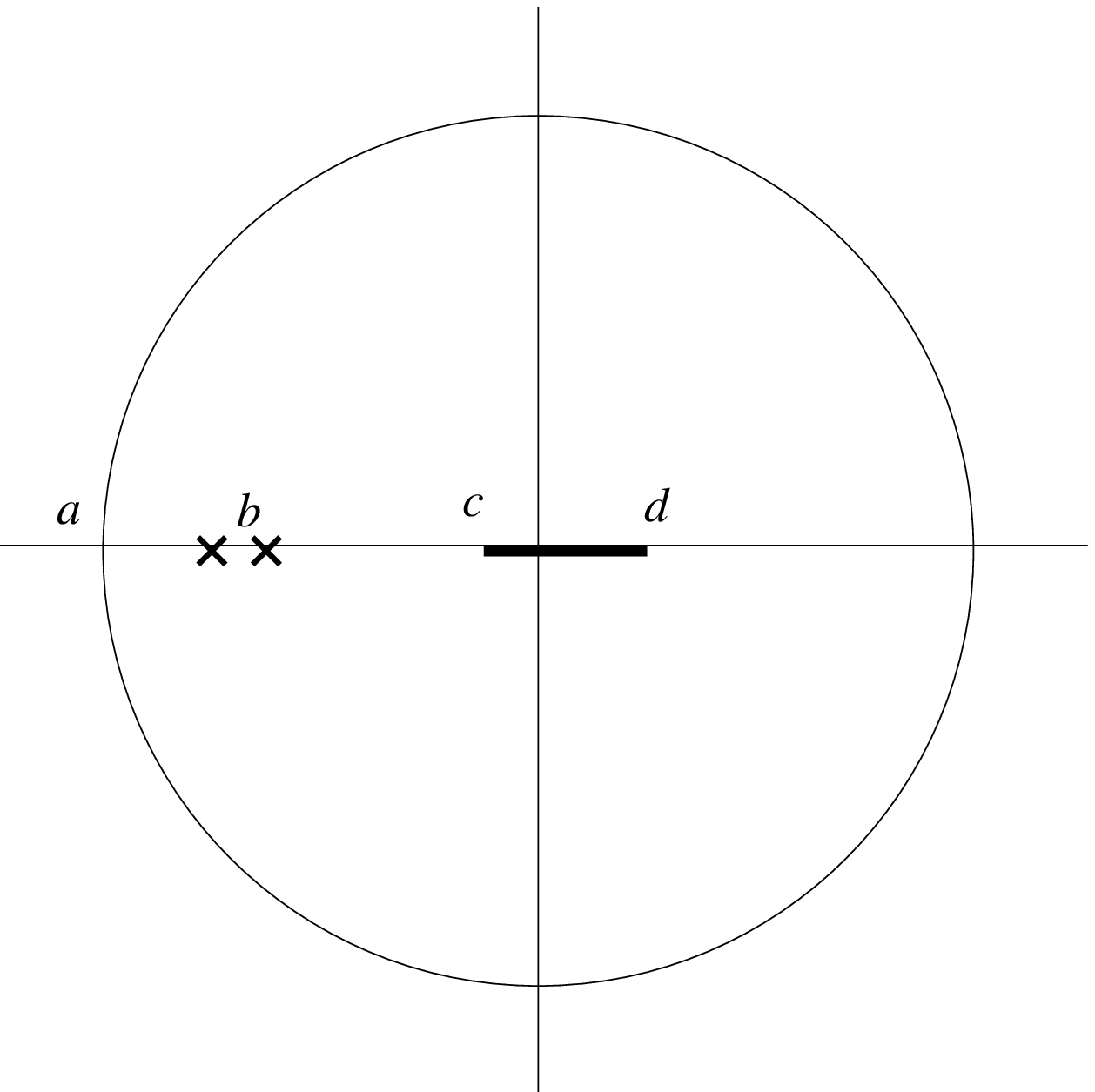}{1}{ Important kinematic points in the $z$-plane
for the related processes (vacuum $\to \overline{M} m$) and $M \to m
\ell \bar \nu$ using the map Eq.~\zdef.  a) ($z=-1$): The branch point
$q^2 = (M+m)^2$, threshold for (vacuum $\to \overline{M} m$).  The two
sides of the branch cut $q^2 > (M+m)^2$ map to the upper and lower
halves of the unit circle, with $q^2 \to +\infty$ corresponding to $z
\to +1$ along the circle.  b) Location of resonant poles below the
(vacuum $\to \overline{M} m$) threshold.  c,d) $z_{\rm min},z_{\rm
max}$ defined in Eq.~\zbounds.  The segment ($z_{\rm min}, z_{\rm
max}$) is the kinematic region for $M \to m \ell \bar \nu$, and $N$ is
chosen in Eq.~\zdef\ so that $z=0$ lies in this interval.  $q^2 \to
-\infty$ corresponds to $z \to +1$ along the real axis.}
where $r \equiv m/M$.  Inasmuch as both $|z_{\rm max}|, \, |z_{\rm
min}| \ll 1$, semileptonic decay possesses a small kinematic expansion
parameter.  One can use the free parameter $N$ to adjust where on the
real axis the interval $(z_{\rm min},z_{\rm max})$ lies, but it is
clear that allowed $z$ values are uniformly smallest when $z_{\rm min}
\leq 0 \leq z_{\rm max}$, for which
\eqn\nbounds{
1 \leq N \leq (M+m)^2/(4 M m) .
}
Choosing $N$ saturates the last remaining degree of freedom allowed by
the Schwarz-Christoffel transformation between the cut plane and the
unit disc.  $N$ may be kinematically interpreted as the value of
$(1+\w)/2$ for which one obtains $z=0$ in Eq.~\zdef.  We explore the
consequences of varying $N$ in the next section.

	Written in terms of $z$, the inequalities from
Eqs.~\disper\--\Xdefs\ now read
\eqn\zbound{
{1\over2\pi i} \int_C {dz \over z} |\phi_i(z) F_i(z) |^2 \leq 1.
}
The contour $C$ is the unit circle.  The weighting functions $\phi_i$
are constructed to be analytic inside the unit circle by multiplying
kinematic factors by functions that are phases on the unit circle. For
example,
\eqn\examples{\eqalign{
{q^2 \over M^2} &\to [(1+r)(1-z) + 2 \sqrt{N r} (1+z)]^2/(1-z)^2 ,\cr
{Q_-\over M^2} &\to {4 r \over(1-z)^2} [(\sqrt{N} -1)z + (\sqrt{N} +
1)]^2 , \cr
{Q_+ \over M^2} &\to 4 r N { (1+z)^2 \over (1-z)^2 }. }}
After some algebra, one obtains the weighting functions
\eqn\phij{\eqalign{
\phi_i = & \sqrt{\kappa } \, 2^{7/2} r^\frac32
         (1+z)^{1 + p} (1-z)^{s +\frac12} N^{\frac34 + \frac{p}2} \cr
          & [(1+r)(1-z) + 2 \sqrt{N r} (1+z)]^{-(s+4)} [(\sqrt{N} -1)z
          + (\sqrt{N} + 1)]^{\frac32 - p},
}}
where $\kappa$, $p$, and $s$ depend on the form factors $F_i$ as
listed in Table~1.
\bigskip
\vbox{\medskip
\hfil\vbox{\offinterlineskip
\hrule
\halign{&\vrule#&\strut\quad\hfil$#$\quad\cr
height2pt&\omit&&\omit&&\omit&&\omit&&\omit&\cr
&i &&F_i&& 1/\kappa\qquad && p && s& \cr
\noalign{\hrule}
&1&& F_0 && 4 \pi \chi^L_V
&& 1 && 0&\cr
&2 && F_1
&& 6 \pi M^2 \chi^{\vphantom{\dagger}}_V &&
0 &&0&\cr
&3&& H_V && 12 \pi M^2\chi^{\vphantom{\dagger}}_V
&& 0&&1 &\cr
&4 && G_0 && 4 \pi \chi^L_A
&&0&&0 &\cr 
&5 && G_1 && 6 \pi M^2\chi^{\vphantom{\dagger}}_A
&&1 &&0 &\cr 
&6 && H_A && 12 \pi M^2\chi^{\vphantom{\dagger}}_A
&&1 &&1 &\cr 
}
\hrule}
\hfil}
\medskip
\INSERTCAP{1}{Factors entering Eq.~\phij\ for the form factors
$F_i$.}

\newsec{Parametrization of Form Factors}

	Equation~\zbound\ constrains the form factors $F_i$ along the
unit circle in the complex $z$ plane, but for semileptonic decay we
are interested in the physical region along the real axis, $z_{\rm
min} \le z \le z_{\rm max}$. If the product $\phi_i(z) F_i(z)$ were
analytic for $ |z| < 1$, one could simply perform a Taylor expansion
about $z=0$.  While the weighting functions $\phi_i$ are analytic
inside the unit circle, the form factors $F_i$ are not.  Contributions
from intermediate states with masses below the $\bar \lb \lc$
threshold lead to cuts and poles in $F_i(z)$ for $ |z| < 1$. While we
expect the contribution from cuts to be unimportant\refs\bglaug, the
more singular nature of poles requires a careful
treatment\refs\spoilers.  The use of Blaschke factors\blasref\ permits
one to eliminate pole contributions\rtc\ given only their masses
$\sqrt{q^2} = m_j$, which are converted to positions $z_j$ inside the
unit circle via Eq.~\zdef.  The Blaschke factors are defined by
\eqn\blaschke{\eqalign{
  P_V =  \prod_{j} {(z -z_j) \over
    (1 - \bar z_j z)} , \cr
  P_A  =  \prod_{j} {(z -z_j) \over
    (1 - \bar z_j z)}, \cr
}}
where $V$ and $A$ indicate the poles appropriate to the polar vector
and axial vector form factors in Eq.~\fdefs, respectively.  Resonances
with spin-parity $J^P= 0^+, 1^-$ couple to form factors of the polar
vector current, while resonances with $J^P= 0^-, 1^+$ couple to the
axial vector current.  The form factors coupled to spin-0 resonances
are exactly those that are multiplied by the Lorentz structure
$q^\mu$, and consequently only appear in the differential width
multiplied by the factor $m_\ell^2$.  It follows that, in the massless
lepton limit, only polar and axial vector resonance masses are needed
in Eq.~\blaschke.

	The usefulness of the Blaschke method lies in the fact that
each factor $(z-z_j)/(1-\bar z_j z)$ serves to eliminate the resonance
pole behavior $1/(z-z_j)$ of the $j$th pole, but is unimodular for
$|z| = 1$.  This holds also for any product of these factors.
Therefore, since each $P_i$ is unimodular on the unit circle, one may
replace $\phi_i F_i$ with $P_i \phi_i F_i$ in the bound Eq.~\zbound\
without changing the result.  Since now both $(P_i F_i)$ and $\phi_i$
are analytic on the unit disc, one can perform a Taylor expansion
about $z=0$,
\eqn\master{
F_i(z) = {1\over P_i(z) \phi_i(z)} \sum_{n=0}^\infty a_n z^n .
}
Substituting this expression into the modified Eq.~\zbound\ gives
\eqn\asum{
\sum_{n=0}^\infty |a_n|^2 \leq 1.
}

	It should be pointed out that one pays a price for these
factors.  The fact that one can eliminate pole contributions without
reference to their residues means that this method must apply equally
well for all allowed values of the pole residues, and therefore the
bound of Eqs.~\master\ and \asum\ is necessarily weaker than the
corresponding bound obtained if one knew anything about the sizes of
the individual residues.  In fact, the bound derived above is not very
restrictive at all, until the constraints supplied by fitting to one
or more experimental data points are imposed.  Then the range of
allowed form factors becomes surprisingly small, as was showed in
Refs.~\refs{\bglapr,\bglaug}.

	Another strength of the parametrization Eq.~\master\ is that,
when $r=m/M$ is not much less than $1$, the scaled kinematic variable
$z$ defined in Eq.~\zdef\ is quite small.  For example, if $N=1$ the
decay $\Lambda_b \to \Lambda_c \ell \bar \nu$ has $z_{\rm min} = 0$
and $z_{\rm max} = 0.049$.  Using Eq.~\asum\ it is clear that the
series in Eq.~\master\ converges quickly, and one can define an
uncertainty from truncating the series after only a few terms.  It
follows that one obtains a functional form for the form factor under
consideration (Eq.~\master) over the whole kinematic range for
semileptonic decay, with a well-controlled and small uncertainty
(because $z$ is small); only the first few parameters $a_n$, each of
which lies in a restricted range (Eq.~\asum), are needed.

	One can do even better by allowing for the parameter $N \neq
1$.  To illustrate this, let us define an approximation $F_i^Q$ by
truncating after the $Q$th term:
\eqn\trunc{
F_i^Q (z) = \frac{1}{P_i(z) \phi_i (z)} \sum^Q_{n=0} a_n z^n .
}
Then the maximum error incurred by truncating after $Q$ terms is just
\eqn\err{\eqalign{
\max | F_i (z) - F^Q_i (z) | & = \max \frac{1}{| P_i (z) \phi_i (z) |}
\sum_{n=Q+1}^\infty | a_n z^n | \cr &
\leq \max \frac{1}{| P_i (z) \phi_i (z)
|} \sqrt{\sum^\infty_{n=Q+1} |a_n|^2} \sqrt{\sum^\infty_{n=Q+1}
z^{2n}} \cr &
< \max_{z \in (z_{\rm min}, z_{\rm max})} \frac{1}{| P_i
(z) \phi_i (z) |} \frac{|z^{Q+1}|}{\sqrt{1-z^2}} ,
}}
where we have used the Schwarz inequality and Eq.~\asum.  Inasmuch as
$| P_i (z) \phi_i (z) | \sqrt{1-z^2}$ varies slowly over the small
interval $z_{\rm min} \leq z \leq z_{\rm max}$, the truncation error
is driven by $|z^{Q+1}|$.  One then sees that the minimum error occurs
for a value of $N$ such that $z_{\rm min} \simeq -z_{\rm max}$.  This
simple observation allows for an improvement over the corresponding
expression in Refs.~\refs{\bglapr,\bglaug}, which used $N=1$ and
$z_{\rm min} = 0$. The interval ($z_{\rm max} - z_{\rm min}$) is
nearly independent of $N$, so $z_{\max}$ for optimal values of $N$ is
approximately half of its value for $N=1$.  Although $z$ is a function
of $N$ through Eq.~\zdef, the Blaschke products $P_i(z)$ for fixed
$q^2$ and pole masses are independent of $N$, while the weighting
functions $\phi(z)$ are only weakly dependent on $N$.  It follows that
for optimal $N$, truncation errors are reduced by a factor of nearly
$2^{Q+1}$, which is appreciable even for small $Q$.

	The reader should be reminded that the functional form
Eq.~\master\ has known, bounded, and small truncation errors,
regardless of the specifics of the experimental data.  A form-factor
expansion in $(\omega -1)$ truncated after linear or quadratic order
introduces theoretical uncertainties that can be substantial\bglaug,
and unlike \master\ is not {\it a priori\/} guaranteed by theory to
fit the data.

\newsec{Heavy Resonances}
     
      To complete the description of the \bcdecay\ form factors
provided by Eq.~\master, one must construct the Blaschke products
$P_i(z)$.  The relevant poles are resonances below the $\bar \Lambda_b
\Lambda_c$ threshold that couple to the $b \to c$ currents.  These are
the polar and axial vector $B_c$ mesons with masses below
$M_{\Lambda_b} + M_{\Lambda_c} \simeq 7.9$ GeV.  Although results on
$B_c$ mesons are still very preliminary\refs\alephbc, their masses are
reliably calculable by interpolating between the charmonium and
bottomonium spectra.  The quark model in a confining potential
predicts a number of relevant $B_c$ polar vector states with the
spectroscopic quantum numbers ${}^3 S_1$ and ${}^3 D_1$.  In the
bottomonium nomenclature, the analogous states are called $\Upsilon
(nS)$ and $\Upsilon (nD)$, respectively. There are also relevant axial
vector states, which divide into nearly degenerate pairs of ${}^1 P_1$
and ${}^3 P_1$ states broken only by spin-orbit and hyperfine
splittings.  The bottomonium analogies are respectively called $h_b
(nP)$ and $\chi_{b1} (nP)$.  The values obtained by various
researchers\refs{\klt{--}\ck} using different approximations in
potential models agree within a few MeV.  We use the values of
Ref.~\eq, supplemented by values for a few higher states above the $B
D^{(*)}$ threshold but below the $\bar \Lambda_b \Lambda_c$
threshold\refs\quigg.  Although the masses of the latter states do not
yet appear in the published literature, their calculation is no more
difficult than that of those already published; however, one should
keep in mind that the simple nonrelativistic bound-state picture for
such $B_c$ resonances is no longer fully justified: The coupling of
open channels to the resonances above the $B D$ threshold
can modify the details of the mass spectrum.  Nevertheless, experience
from the $J/\psi$ and $\Upsilon$ systems shows that such calculations
give a reasonable account of masses above the $\bar D D$ and $\bar B
B$ thresholds.  We present the $B_c$ pole mass values used in Table~2.
\bigskip
\vbox{\medskip
\hfil\vbox{\offinterlineskip
\hrule
\halign{&\vrule#&\strut\quad\hfil$#$\quad\cr
height2pt&\omit&&\omit&&\omit&\cr
& {\rm Type} && {\rm Source} && {\rm Masses} \; {\rm (GeV)} \qquad
& \cr \noalign{\hrule}
& {\rm Vector} && {\rm Ref.~\eq} && 6.337, \; 6.899, \; 7.012, \;
7.280 &\cr
&  && {\rm Ref.~\quigg}
&& 7.350, \; 7.594, \; 7.646, \; 7.872 \; 7.913 &\cr
\noalign{\hrule}
& {\rm Axial} && {\rm Ref.~\eq} && 6.730, \; 6.736, \; 7.135, \; 7.142
& \cr
&  && {\rm Ref.~\quigg}
&& 7.470, \; 7.470, \; 7.757, \; 7.757 &\cr
}
\hrule}
\hfil}
\medskip
\INSERTCAP{2}{Assumed $B_c$ pole masses used in this work.  A complete
set of masses for either parity consists of the values from Ref.~\eq,
together with the newly-computed values from Quigg\quigg.}

	The truncation errors on our parametrizations depend on the
Blaschke factors only through the endpoint values $P(z_{\rm min})$ and
$P(z_{\rm max})$, because empirically the maximum truncation errors
from Eq.~\err\ occur at either $z=z_{\rm min}$ or $z_{\rm max}$.
These values of $P$ do not depend on $N$, but they do depend on the
masses of states between the $B D$ and $\bar \Lambda_b \Lambda_c$
thresholds.  We present the truncation errors for $Q =2$ (a
two-parameter description if given the overall normalization, or three
parameters otherwise) in Table~3. Truncation errors for $Q=3$ are
roughly a factor of 40 times smaller because of the additional
suppression of about $z_{\rm max}^{N=1}/2$ described in Sec.\ 3. To
compare these absolute truncation errors to the expected size of the
form factors, note that heavy quark symmetry predicts the
normalization of $F_1$ and $G_1$ at zero recoil to be unity, while the
other form factors are ${\cal O} (1/m_c)$ in this limit\refs\iwbary.
Empirically, the value $N = 1.09$ minimizes the truncation errors,
although this number may change slightly depending on the actual
values of $P(z_{\rm min})$ and $P(z_{\rm max})$.  These truncation
errors are substantially smaller than those for $N=1$, which is $28\%$
for $F_1$, for example.  Randomly altering the resonance masses by
20--50 MeV alters the truncation errors presented in Table~3 by less
than one part in ten in each case, indicating the insensitivity of the
errors to the particular assumptions of the potential model used to
compute $B_c$ masses.
\bigskip
\vbox{\medskip
\hfil\vbox{\offinterlineskip
\hrule
\halign{&\vrule#&\strut\quad\hfil$#$\quad\cr
height2pt&\omit&&\omit&&\omit&\cr
& F_i && {\rm (Max.} \; {\rm Trunc.}
\; {\rm error)} \cdot 10^2 && N_{\rm ideal} &\cr
\noalign{\hrule}
& F_0 && 4.0 \qquad\qquad && 1.088 &\cr
& F_1 && 5.1 \qquad\qquad && 1.088 &\cr
& H_V && 20.0 \qquad\qquad && 1.089 &\cr
& G_0 && 4.3 \qquad\qquad && 1.089 &\cr
& G_1 && 6.6 \qquad\qquad && 1.088 &\cr
& H_A && 25.7 \qquad\qquad && 1.089 &\cr
}
\hrule}
\hfil}
\medskip
\INSERTCAP{3}{Truncation errors from Eqs.~\trunc\ and \err\ of the
\bcdecay\ form factors for the pole masses given in Table~2 with
$Q=2$.  $N_{\rm ideal}$ is the value of $N$ for which these errors are
minimized.  Including all other known corrections multiplies the
truncation errors by the coefficient $B \leq 1.4$ described in
the text.}

	Finally, we briefly commented in Sec.\ 3 that the effect of
branch cuts due to nonresonant on-shell intermediate states connecting
the vacuum to $\bar \Lambda_b \Lambda_c$ are not numerically
significant, just as in $\bar B \to D^{(*)} \ell \bar \nu$.  In
Ref.~\bglaug\ their effects were accommodated by relaxing the bound
\asum\ to $\Sigma |a_n|^2 \le B^2$, with $B \le 1.05$.  In the present
case, however, the $\bar \Lambda_b \Lambda_c$ threshold is higher than
the $B D^{(*)}$ threshold, and so many more cuts are possible.  To
argue that their total effect is still negligible requires a more
detailed analysis than appears in \bglaug.

	We begin with the observation that multiparticle
intermediate states connect the current $J^\mu$ to $\bar \Lambda_b
\Lambda_c$, and thus appear as loop diagrams.  Each additional
intermediate state introduces a loop and large-$N_{\rm color}$ factor
of 1/( 16$\pi^2 N_c^{1/2})$, so we limit ourselves to two-particle
intermediates. In reference \bglaug, we modeled the contribution of
such a state with a square-root cut, as follows. If the combined mass
of the intermediate particles corresponds to $z = z_{\rm cut}$, the
cut contribution to $B$ is
\eqn\delB{
\delta B = \Biggl[ {1\over{2\pi}}\int_0^{2\pi} d\theta\, 
                      |g_{\rm cut}(z)\phi(z)|^2~ \Biggr]^{1/2},
}
where $z=e^{i\theta}$, $\phi(z)$ is the 
weighting function, and
\eqn\cutforminz{
g_{\rm cut}(z)={ 4c\sqrt{r} \over M_{\Lambda_b} }
                      \left({\sqrt{(z-z_{\rm cut})(1-zz_{\rm cut})}
                      \over (1-z)(1-z_{\rm cut})} - {1+z\over 2 (1-z)}
                      \right) }
models the cut.

	The coupling constant $c$ may be written as $c = \hat f \hat
g/8\pi$, where $ \hat f$ is the coupling between the current and the
two-particle intermediate state, $ \hat g$ is the overlap between
these states and $\bar \Lambda_b \Lambda_c$, and the $8\pi$ comes from
the loop.

	Two-particle intermediate states with strangeness are OZI
suppressed, so the only kinematically allowed possibilities are $B-D$
resonances and states with one $B_c$ resonance and one light
unflavored meson.  We may further limit our consideration to $S$-waves
only, since higher partial amplitudes are suppressed by powers of the
available three-momentum, ${\cal O} (\Lambda_{\rm QCD})$ because we
are in the resonance region, divided by the center-of-mass energy,
${\cal O} (M_B + M_D)$.  Furthermore, two-particle intermediates in
which one of the particles has strong isospin (such as $B_c \pi$) are
heavily suppressed by isospin-breaking coefficients, and will be
ignored.

	With these restrictions, the decay $\bar B \to D^{*} \ell \bar
\nu$ admits only two very short cuts in the $1^-$ channel ($B_c
({}^3P_1) \eta$ and $B_c({}^{1 \!}P_1) \eta$), with $z_{\rm cut}
\approx -0.88$, and three in the $1^+$ channel, with the lowest being
$B_c^* \eta$ with $z_{\rm cut} = -0.48$. There are anomalous cuts
starting at $z \approx -0.35$, but they are proportional to a
$B^*$-$B$-$\pi$ coupling $g^2$ that is probably quite small\dec. $\bar
B \to D \ell \bar \nu$ has no cuts in the relevant $1^-$ channel,
anomalous or otherwise. These branch cuts lead to even smaller
uncertainties for $\bar B \to D^{(*)} \ell \bar \nu$ than anticipated
in Ref.~\bglaug.  On the other hand, once the threshold is raised to
the $\Lambda_b + \Lambda_c$ mass, many more cuts occur.  We count 32
in the $1^-$ channel and 28 in the $1^+$ channel, including such
exotic combinations as $B_c (2 \, {}^{3 \!}P_0 (6700)) + h_1 (1170)$.
To justify our neglect of so many cuts requires a more careful study.

	The great majority of cuts in either channel arise from a
$B_c$ resonance and a light unflavored meson.  To get a rough estimate
of their combined contribution, we take them all to have the same
coupling $c$, compute the contribution to $B$ of each state
separately, and add them in quadrature (to simulate random phases
between the various cuts).  We find that $B$ is only increased to $1 +
0.5 c$ for the $1^-$ channel and $1+ 0.6 c$ for the $1^+$
channel. Furthermore, we expect $c = \hat f \hat g/8\pi$ to be very
small. To judge the typical size of $\hat f$, for example, we can
extract the coupling $\hat f_{\psi\eta}$ of a vector current to the
$J/\psi$-$\eta$ state from charmonium radiative decay data.  Using a
phenomenological Lagrangian term
\eqn\lag{\eqalign{ \delta {\cal L} = i e \hat f_{\psi\eta} M_\psi
\psi^\mu A_\mu \eta}}
gives a coupling $\hat f_{\psi\eta} < 10^{-2}$, while bottomonium data
gives an even smaller bound on $\hat f_{\Upsilon\eta}$.  Taking the
interpolated value of $\hat f_{B_c\eta}$ as representative, we
conclude that cuts involving $B_c$ resonances give negligible
contribution to $B$.

	On the other hand, the intermediate states consisting of a
$B$-$D$ pair may have much larger couplings, with $\hat f\hat g$
conceivably of order one, but the cuts begin much closer to the $\bar
\Lambda_b \Lambda_c$ threshold and are fewer in number (six for the
$1^-$ channel, with minimum $z_{\rm cut} = -0.55$, three for the $1^+$
channel, with minimum $z_{\rm cut} = -0.39$).  Added in quadrature,
the bounding parameter $B$ in the vector (axial) channel receives a
$0.12 c$ ($0.18 c$) correction.  However, since our uncertainty
estimates to obtain $B$\bglaug\ are added in quadrature, the
conservative total correction to Eq.~\asum\ given by $B \le 1.4$
remains unchanged in this work, even if $c = {\cal O} (1)$.  In
addition to multi-particle cuts, these uncertainty estimates include
corrections to the heavy quark limit, choices of quark mass values,
and ${\cal O} (\alpha_s)$ perturbative corrections, as well as others.
A value of $B = 1.4$ has the effect of increasing all of our
truncation errors by a factor of $1.4$; such increases on a
few-percent truncation error do not hamper our conclusions.

\newsec{Implications for Meson Decays}

	The freedom to choose $N$ in Eq.~\zdef\ has significant
implications for semileptonic meson decays like $\bar B \to D^{(*)}
\ell \bar \nu$ and $\bar B \to \pi \ell \bar \nu$.  For $\bar B \to D
\ell \bar \nu$ and $\bar B \to D^* \ell \bar \nu$, the form factors
defined by
\eqn\fdefsbd{\eqalign{
  \bra{D^*(p',\eps)} V^\mu \ket{\bar B(p)} =& \, i g \epsilon^{\mu
    \alpha \beta \gamma} \epsilon_\alpha^* \, p_{\beta}' \, p_{\gamma}
    \cr
  \bra{D^*(p',\eps)} A^\mu \ket{\bar B(p)} =&
   \, f_0 \epsilon^{*\mu} + (\epsilon^{*} \! \cdot p)
   [a_+ (p + p')^\mu + a_- (p - p')^\mu] \cr
   \bra{D(p')} V^\mu \ket{\bar B(p)} =& \, f_+ (p + p')^\mu + f_-
   (p - p')^\mu \cr }}
and
\eqn\fcdef{
F_1= {1\over m} \left[2q^2 k^2 a_+ - \frac12 (q^2-M^2+m^2)f_0
\right] ,
}
where now $M = M_B$ and $m = M_{D^{(*)}}$, have weighting functions
\eqn\phibd{\eqalign{
\phi_i = & M^{2-s} \sqrt{\kappa n_f} \, 2^{2 + p/2}
	 [r(1+z)]^{1 + p\over 2}
         (1-z)^{s -\frac32} N^{\frac12 + \frac{p}4} \cr & [(1+r)(1-z)
         + 2 \sqrt{N r} (1+z)]^{-(s+p)} [(\sqrt{N} -1)z + (\sqrt{N} +
         1)]^{\frac{p}2},
}}
where $n_f$ is an isospin Clebsch-Gordan factor that is 2 for $\bar B
\to D^{(*)} \ell \bar \nu$ and $\frac{3}{2}$ for $\bar B \to \pi \ell
\bar \nu$, $r \equiv m/M$, and the values of $\kappa$, $p$, and $s$
are given in Table~4.  The parameters $p$ and $s$ here do not have
precisely the same origin as those in Eq.~\phij.  Equation~\phibd\
agrees with Ref.~\bglaug\ in the limit $N \to 1$.
\bigskip
\vbox{\medskip
\hfil\vbox{\offinterlineskip
\hrule
\halign{&\vrule#&\strut\quad\hfil$#$\quad\cr
height2pt&\omit&&\omit&&\omit&&\omit&&\omit&\cr
&i &&F_i&& 1/\kappa\qquad && p && s& \cr
\noalign{\hrule}
&0&& f_0 && 12 \pi M^2\chi^{\vphantom{\dagger}}_A
&& 1 && 3 &\cr
&1 && F_1
&& 24 \pi M^2 \chi^{\vphantom{\dagger}}_A &&
1 &&4&\cr
&2&& g && 12 \pi M^2\chi^{\vphantom{\dagger}}_V
&& 3 &&1 &\cr
&3 && f_+ && 6 \pi M^2\chi^{\vphantom{\dagger}}_V
&& 3 &&2 &\cr }
\hrule}
\hfil}
\medskip
\INSERTCAP{4}{Factors entering Eq.~\phibd\ for the meson form factors
$F_i$.}
{}From the discussion in Sec.\ 3, one expects that by choosing the
optimal value between $N=1.10$ and 1.12 (the exact value depends upon
the form factor), the truncation errors for a three-parameter
description should be reduced by roughly a factor of 8 over the $N=1$
values in Ref.~\bglaug, while the truncation errors for a
two-parameter fit should be reduced by a factor of about 4.  {}From
direct calculation, these numbers are 6.1--7.2 and 3.3--3.7, in accord
with expectations.  Since one of the parameters is determined from the
normalization of the form factor at zero recoil by heavy quark
symmetry, the $N \simeq 1.1$ choice provides a {\it one}-parameter
description of each of the form factors with truncation errors
(relative to the normalization at zero recoil) of 6.4\%, 5.0\%, 2.9\%,
and 30.9\% for $f_+$, $f_0$, $g$, and $F_1$, respectively.  For the
form factor $g$, for example, the conclusion is that using the
normalization at zero recoil (which determines $a_0$) plus the slope
at the same point (which determines $a_1$), one obtains the shape of
$g$ over the entire kinematic range with a theoretical error of no
more than 2.9\%.  Uncertainties like corrections to heavy quark
symmetry, perturbative QCD corrections, and so on are expected to
increase this theoretical error to no more than 4.1\% and possibly
much less\bglaug.  For example, the truncation error vanishes at zero
recoil, where the normalization is known.

	We exhibit the possible shapes of $g(\w)/g(1)$ in Fig.~2.  The
lowest curve corresponds to the saturation of Eq. \asum\ by $a_1 =
-1$, while the top curve corresponds to $a_1 = 0.115$, the largest
value allowed by the the Bjorken inequality\refs\bjbd\ on the slope of
the Isgur-Wise function improved by ${\cal O} (\alpha_s)$
corrections\refs\bgm. Intermediate curves correspond to equally spaced
values of $a_1$.

	The form factor $f_+^\pi$ defined by replacing the $D$ meson
in Eq.~\fdefsbd\ with a pion has the same weighting function $\phi_+$
as $f_+^D$, but with the $D$ mass replaced with the pion mass and
$\chi_V = 9.5 \cdot 10^{-3}/m_b^2$. The Blaschke factor for the pion
form factor is $P^\pi = (z - z_*)/(1 - z z_*)$, with $z_*$
corresponding to the $B^*$ mass, the sole resonance below threshold.
One might expect a suitable choice of $N$ to lead to $|z_{min}|,
|z_{max}|$ of about half the $N=1$ value of $z_{max}^{N=1} \simeq
0.5$.  Unfortunately, one cannot ignore the variation of $P(z) \phi(z)
\sqrt{1-z^2}$ over the kinematic range, so the reduction in truncation
errors is less spectacular than in the $B \to D$ system. It is
nevertheless a significant improvement: Where for $N=1$, a
four-parameter description implied an absolute truncation error of
$1.35$, for $N=2.1$ the absolute truncation error is $0.37$.  A
five-parameter description yields, for $N=2.27$, a $0.11$ truncation
error.  Even for a three-parameter description, the absolute
truncation error at $N=1.86$ is 1.19 over the large kinematic range of
$\bar B \to \pi \ell \bar \nu$.  In typical models, $f_+^\pi(q^2)$
varies from around $0.3$ at $q^2=0$\refs\models\ to order $10$ at $q^2
= (M_B - m_\pi)^2$\refs\dec.  Other models may give smaller values of
$f_+^\pi (q^2_{\rm max})$ or fall off rapidly away from $q^2_{\rm
max}$, and then it is useful and straightforward to choose $N$ in
order to minimize the truncation errors {\it relative\/} to particular
points fixed by a model, instead of the absolute errors quoted above.

\newsec{Conclusions}

	The analysis of dispersion relations and analyticity
properties of strong-interaction amplitudes, once ubiquitous in
particle theory, can still yield a surprising amount of information
about heavy hadron transitions. For the semileptonic decays considered
in this paper, this stems from a two-part procedure: First, the
perturbative calculation of a two-point function, performed in an
unphysical kinematic regime where the calculation is reliable, is
connected by crossing symmetry and a dispersion relation to the QCD
form factors of interest. Second, some complex analysis consisting of
a conformal transformation, a multiplication by Blaschke functions,
and a Taylor expansion, produces a simple parametrization for the form
factors in the physical region.

	In the current work we have presented an improvement of the
conformal transformation that decreases the number of parameters
necessary for an accurate description of the form factors over a given
kinematic range. We have computed the explicit parametrizations for
the form factors of the decay \bcdecay, and have recomputed, using the
new conformal transformation, the parametrizations of the mesonic
decay form factors for $\bar B \to D^{(*)} \ell \bar \nu$ and $\bar B
\to \pi \ell \bar \nu$.

	In particular, for a two-parameter description of \bcdecay\
form factors (utilizing heavy quark symmetry for the normalization), we
find truncation errors of 4--7\% for the form factors associated with
transversely polarized lepton pairs, and errors of 20--26\% for the
longitudinally polarized combinations.  These numbers would have been
much larger without the technical improvement to the conformal map
Eq.~\zdef, owing to the large number of $B_c$ resonance masses below
the $\bar \Lambda_b \Lambda_c$ threshold.  The truncation errors are
seen to be quite insensitive to the exact locations of the $B_c$
poles, so our results do not depend on the detailed assumptions of a
particular potential model calculation.

	Constraints for the decays $\bar B \to D^{(*)} \ell \bar \nu$,
which were already quite restrictive from earlier work, become more so
using this technical improvement.  In particular, we showed that one
can obtain a one-parameter fit to all of these form factors except
$F_1$ good to $3$--$6\%$, using the normalization from heavy quark
symmetry.  A description of the form factor $f_+$ of $\bar B \to \pi
\ell \bar \nu$ good over the entire kinematic range to an absolute
uncertainty of 0.37 requires only four parameters. Relative to the
expected size of $f_+$, this represents a small fractional uncertainty
in the pole-dominated region.  In addition, if the normalization of
$f_+$ at more than one kinematic point is known, the envelope of
allowed parametrizations becomes quite restricted, even for small
$q^2$\dec.

	These conclusions follow only from very general properties of
QCD and $S$-matrix amplitudes. They do not rely on such potentially
useful information as the contribution of exited states above
threshold to the dispersion relation, or the actual contribution
(residues) of heavy resonances below threshold to singularities of the
form factors. One might therefore expect only very weak constraints to
arise; instead one is rewarded with strong restrictions on the shape
of the form factors. In this paper, our treatment has been somewhat
mathematical. A physical understanding of why the restrictions are so
strong, and what role the small but apparently natural variable $z$
plays, would be of considerable interest.  Further detailed
computations compatible with the QCD dispersion constraints may yield
restrictions that are better still.

\vskip1.2cm
{\it Acknowledgments}
\hfil\break
We would like to thank Ben Grinstein and Aneesh Manohar for valuable
conversations, Vivek Sharma for a discussion of current experimental
information, and especially Chris Quigg, who computed a number of
$B_c$ mass eigenvalues used in this paper.  This work is supported in
part by the Department of Energy under contract DOE--FG03--90ER40546.

\vfill\eject

\vfill\eject \listrefs\eject

{\bf Figure Captions}

Figure 1.  Important kinematic points in the $z$-plane for the related
processes (vacuum $\to \overline{M} m$) and $M \to m \ell \bar \nu$
using the map Eq.~\zdef.  a) ($z=-1$): The branch point $q^2 =
(M+m)^2$, threshold for (vacuum $\to \overline{M} m$).  The two sides
of the branch cut $q^2 > (M+m)^2$ map to the upper and lower halves of
the unit circle, with $q^2 \to +\infty$ corresponding to $z \to +1$
along the circle.  b) Location of resonant poles below the (vacuum
$\to \overline{M} m$) threshold.  c,d) $z_{\rm min},z_{\rm max}$
defined in Eq.~\zbounds.  The segment ($z_{\rm min}, z_{\rm max}$) is
the kinematic region for $M \to m \ell \bar \nu$, and $N$ is chosen in
Eq.~\zdef\ so that $z=0$ lies in this interval.  $q^2 \to -\infty$
corresponds to $z \to +1$ along the real axis.

Figure 2.  The one-parameter description of the $\bar B \to D^* \ell
\bar \nu$ form factor $g(\w)/g(1)$, plotted as a function of velocity
transfer $\w$, for a set of values of the parameter $a_1$. The lowest
curve corresponds to the saturation of Eq. \asum\ by $a_1 = -1$, while
the top curve corresponds to $a_1 = 0.115$, the largest value allowed
by the the $\alpha_s-$improved Bjorken inequality\refs{\bjbd,\bgm}.
Intermediate curves correspond to equally spaced values of $a_1$.
\bye